\documentclass[12pt,a4paper,final]{iopart}

\usepackage{iopams}
\usepackage{graphicx}
\usepackage[breaklinks=true,colorlinks=true,linkcolor=blue,urlcolor=blue,citecolor=red]{hyperref}
\usepackage{verbatim}
\usepackage{cite}
\usepackage{multirow}
\usepackage{url}
\usepackage{ulem}

\begin{document}

\title[local vibrational modes]{Local vibrational mode of impurity in a monatomic linear chain under open and periodic boundary conditions}

\author{Qiang Luo}

\address{ Department of Physics, Renmin University of China, Beijing, 100872, China}
\ead{qiangluo@ruc.edu.cn}

\begin{abstract}
In this paper, we revisit the lattice vibration of one-dimensional monatomic linear chain under open and periodic boundary conditions, and give the exact conditions for the emergence of the local vibration mode when one of the atoms is replaced by an impurity. Our motivation is twofold. Firstly, in deriving the dispersion relation of the atoms, periodic boundary condition is overwhelmingly utilized while open boundary condition is seldomly referred. Therefore we manage to obtain the dispersion relation under both boundary conditions simultaneously by Molinari formula. Secondly, in the presence of impurity, local vibration mode can emerge as long as the mass of the impurity $m'$ is smaller than the mass of the perfect atom $m$ to certain degree, which can be measured by the mass ratio $\delta=\frac{m-m'}{m}$. At periodic boundary condition, the critical mass ratio is $0$ or $\frac{1}{N}$, depending on whether the length $N$ of the chain is even or odd. At open boundary condition, the critical mass ratio is $\frac{N}{2N-1}$ if the impurity locates at the end of the chain, while it is $\frac{N}{(2N_l+1)(2N_r+1)}$ with $N_l$ and $N_r$ be the number of atoms at the left and right hand sides of the impurity if the impurity locates at the middle.

\vbox{}

\end{abstract}
%\submitto{\EJP}
%\maketitle
\section{Introduction}
The physical properties of low dimensional materials have attracted attentions of physicists especially at the background of the synthesis of carbon nanotube and graphene\cite{TS-2014}, therefore it is the interest of us to revisit the enduring topic of lattice vibration of one-dimensional (1D) linear chain. In the perfect crystal lattice, the atoms vibrate along their own equilibrium positions incessantly, and the interaction among them makes those local vibrations connect, which, of course, produce the so-called lattice wave. Impurities, however, are known to affect the vibrational properties of crystals by modifying the distribution of normal modes and thus may induce some unusual vibrational modes.

The 1D monatomic linear chain and diatomic linear chain are workhorses for the illustration of lattice wave as well as acoustic and optical branches. In the case of nearest neighbor interaction and harmonic approximation, the equation of motion of each individual atom is governed by the second law of Newton. For the finite chain we must specific how the atoms at the two endpoints are to be described to avoid the possible ambiguousness. It is generally believed that the boundary condition is not of significant importance in the infinite chain since it fails to alter the final results essentially. The standard textbooks on solid state physics\cite{Kittel-1996,AM-1976,GP-2005}, without exception, will tell us to avoid the cumbersome open boundary condition (OBC) and instead utilize the periodic boundary condition (PBC), based on which only one or two algebraic equations are needed to obtain the dispersion relation of monatomic chain or diatomic chain. Although in most cases the buck properties of \textit{infinite} chain do not rely on the boundary condition, it does make a difference if the length of chain is \textit{finite}. Furthermore, the PBC prevents us from the study of surface effects. The overwhelming utilization of PBC in the derivation of the dispersion relations in most available materials will certainly mislead the students, who may underestimate the true effect of the boundary condition.

Historically, the vibration frequencies of 1D lattice with PBC and OBC were studied by Max Born in a series of papers already, though the solution to the OBC case is not really exact. Later J. O. Halford\cite{Halford-1951} and Richard F. Wallis\cite{Wallis-1957} rederived the dispersion relation with OBC with the help of the determinants of two special cases summarized by D. E. Rutherford\cite{Rutherford-1947,Rutherford-1952}. To explain the differences between PBC and OBC, we shall take the local vibration due to the impurity as an example. It is expected that the boundary condition will alter the condition for the emergence of the local vibration.

For simplicity we concentrate on the 1D monatomic linear chain, while the 1D diatomic linear chain is left for the motivated readers. Our paper is organized as follow. In section \ref{sec1Dmonatomicdispersion} we introduce the Molinari formula, and through which we derive the dispersion relation of 1D monatomic chain under open and periodic boundary conditions. In section \ref{sec1Dmonatomicvibrationmode} we give the exact conditions for the emergence of the local vibration mode of impurity at different positions of the linear chain and at different boundary conditions. The local vibration frequency and amplitude are presented at the subsequent section \ref{sec1Dmonatomicvibrationfrequency}. Section \ref{secConclusion} is devoted to our conclusions.
%=====================================================================================================================================================
\section{The dispersion relation of 1D monatomic chain}\label{sec1Dmonatomicdispersion}
We shall review the dispersion relation of 1D monatomic chain where only one atom per primitive cell of lattice constant $a$ and force constant $\beta$, formed by $N$ atoms of mass $m$. The $n$-th atom oscillates around its equilibrium position $na$ with the displacement $u_n$. With the simple harmonic approximation and the nearest neighbor approximation the motion of the $n$-th atom can be given by Newton's law
\begin{equation}\label{NewtonLaw}
m\frac{d^2u_n}{dt^2} = -\beta\left(-u_{n-1}+2u_n-u_{n+1}\right)
\end{equation}
with $n=1,2,\cdots,N$. The PBC which means that when we come to the end of the atomic chain we assume that the chain identically repeats itself is used, \textit{i.e.} $u_0=u_N$ and $u_{N+1}=u_1$. In the OBC, however, the equation of motions of the two atoms siting at the boundary are
\begin{equation}\label{NewtonLawEndPoint}
\cases
{m\frac{d^2u_1}{dt^2} = -\beta\left(u_{1}-u_{2}\right)\\
 m\frac{d^2u_N}{dt^2} = -\beta\left(-u_{N-1}+u_{N}\right)}.
\end{equation}

The set of discrete coupled differential equations can be solved by means of travelling waves with wavevector $q$ and frequency $\omega$, of the form
\begin{equation}\label{travellingWave}
u_{n}(t) = \bar{u}_q(n)e^{-i\omega t}.
\end{equation}
Using the equations above we therefore arrive at the secular equation of the monatomic chain
\begin{equation}\label{secularEq}
M^{(l)}\bar{U}=0
\end{equation}
where $\bar{U}=\left(\bar{u}_{q}(1),\bar{u}_{q}(2),\cdots,\bar{u}_{q}(N)\right)^{T}$, and
\begin{equation}\label{matrixM}
M^{(l)}=
\left(
  \begin{array}{ccccc}
    1+l-E & -1      &        &        & -l       \\
    -1    & 2-E     &     -1 &        &          \\
          & \ddots  & \ddots & \ddots &          \\
          &         &     -1 & 2-E    & -1       \\
    -l    &         &        & -1     & 1+l-E    \\
  \end{array}
\right)
\end{equation}
with $E=m\omega^2/\beta$, and $l=0$ and 1 for OBC and PBC, respectively. The non-trivial solution to \eref{secularEq} exists unless the determinant of the coefficient matrix \eref{matrixM} equals to zero, \textit{i.e.} $\vert M^{(l)}\vert=0$ . The determinant of \eref{matrixM}, a tridiagonal matrix, can be calculated by various methods, such as Chebyshev polynomial method~(see Appendix A) and Molinari formula method~(see Appendix B).

Tridiagonal matrix and its generalized form, which are of analytical and numerical importance, have made an extensive and profound influence on physics due to their unique mathematical structures. On the one hand, the tight-binding approximation and the nearest neighbor interaction approximation allow us to transform the Hamiltonians of numerous ideal models to tridiagonal matrices. On the other hand, it is possible to reduce the Hamiltonian of any physical problem to a tridiagonal matrix by virtue of some standard numerical algorithms such as Lanczos method. We shall mention here that a determinant should be in hand to obtain the quantum fluctuation factor of the harmonic oscillator in the framework of Feynman's path integral\cite{Hira-2013}.

Abundant methods have been discovered in view of the widespread applications of the determinant of these tridiagonal matrices. Rutherfold\cite{Rutherford-1947,Rutherford-1952} had already summarized many useful conclusions relating to the determinant that had made their appearances in physics and chemistry around the middle of the last century. The simple and explicit analytic expression which is used to calculate the determinant of arbitrary tridiagonal matrices, however, is absent until about two decades ago when Molinari finally found such a splendid formula\cite{Molinari-1997,Molinari-2008}. Molinari formula is powerful in that it can handle the tridiagonal matrices whose diagonal elements and/or offdiagonal elements are not necessarily the same. This feature makes it possible for us to settle impurity problem which will come in the subsequent sections. From the Molinari formula we find that
\begin{equation}\label{detM}
\cases
{\vert M^{(1)}\vert = 2\left(\cos N\tau-1\right)\\
 \vert M^{(0)}\vert = -2\tan\frac{\tau}{2}\sin N\tau}
\end{equation}
where the dimensionless parameter $\tau\equiv qa$ satisfies $2-E=2\cos\tau$. The dispersion, therefore, should be
\begin{equation}\label{dispersionRelation}
E^{(l)} = 2\left(1-\cos qa\right).
\end{equation}

\begin{figure}[!h] %[t]
\centerline{\includegraphics[width=9.0cm,height=5.0cm]{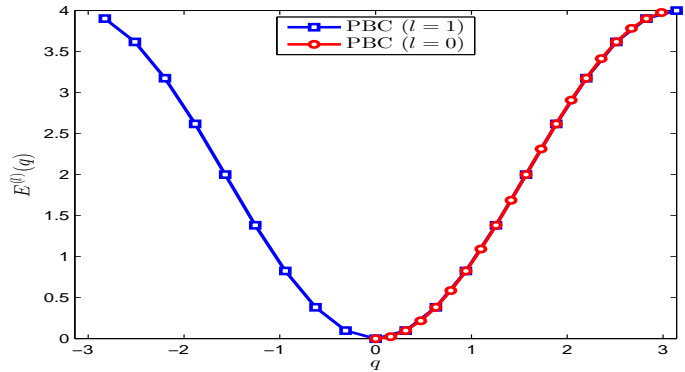}\hspace{4mm}}
%\centerline{\includegraphics[height=5.0cm,width=16.5cm]{fig1.eps}\hspace{4mm}}
%\begin{minipage}[b]{9cm}
\caption{The dispersion relation of 1D monatomic chain with $N=20$ under open~(red symbols) and periodic~(blue symbols) boundary conditions.}\label{FIG-MCDisp}
%\end{minipage}
\end{figure}

Since the dispersion relation is at hand, we can turn to the possible wavevectors which lie in the first Brillouin zone(FBZ), \textit{i.e.} the Wigner-Seitz unit cell of the reciprocal lattice. In the FBZ, {\color{blue}{the number of possible vibration wavevectors equals to that of the number of degree of freedom. Besides, the determinant of $M^{(l)}$~($l=0,1$) in \eref{detM} will be zero on condition that $\cos N\tau=1$ ($l=1$) or $\sin N\tau=0$ ($l=0$), which fulfills the traditional analysis that $e^{iNqa}=1$. For the PBC~($l=1$), the wavevectors satisfy the relation $Nq_{1}a=2\pi h_{1}$, while for the OBC~($l=0$) it is $Nq_{0}a=\pi h_{0}$ with both $h_1$ and $h_0$ are some integers\cite{QLUO-Note}. Therefore, the wavevectors in the FBZ should be
\begin{equation}\label{WaveVectorFBZOBCPBC}
q_{l}=
   \cases
   {\frac{2\pi}{Na}h_{1},              &\textrm{PBC} \\
    \frac{ \pi}{Na}h_{0},              &\textrm{OBC}}
\end{equation}
with $h_{1} = -\frac{N}{2}+1, -\frac{N}{2}+2, \cdots, \frac{N}{2}$ and $h_{0} = 0, 1, \cdots, N-1$, a set that totally has $N$ different values\cite{Yueh-2005}, and the dispersion relation is shown in \fref{FIG-MCDisp}\cite{QLUO-Note}. }}

%the wavevectors satisfy the condition  $-\pi<qa\leq\pi$, or $-\frac{\pi}{a}<q\leq\frac{\pi}{a}$ more precisely. Besides, the determinant of $M^{(l)}$~($l=0,1$) in \eref{detM} will be zero on condition that $\cos N\tau=1$ ($l=1$) or $\sin N\tau=0$ ($l=0$), which is in accordance with the traditional analysis that $e^{iNqa}=1$. Anyway, $Nqa$ can be any number that is the integer multiplying $2\pi$, \textit{i.e.} $Nqa=2\pi h$. So the wavevectors in the first BZ should be
%\begin{equation}\label{WaveVectorFBZ}
%q = \frac{2\pi}{Na}h,
%\end{equation}
%where $h = -\frac{N}{2}+1, -\frac{N}{2}+2, \cdots, \frac{N}{2}$, a set that totally has $N$ different values.

\section{Local vibrational mode of impurity of 1D monatomic chain}\label{sec1Dmonatomicvibrationmode}
We now turn to the situation where one of the atoms in the chain is replaced by an impurity with mass $m'$. It is generally believed that the mass and the position of the impurity will influence the local vibration, the vibration frequency and amplitude for instance. We define the mass ratio
\begin{equation}\label{deltam}
\delta=\frac{m-m'}{m}
\end{equation}
as the measurement of the relative mass of impurity. Since it is known that the localized vibrational exists only for the light mass impurity\cite{AD-2000}, in the subsequent sections we will discuss the case $0<\delta<1$ only under the PBC and OBC, respectively. {\color{blue}{Further, the maximum vibration frequency~(or energy) of the atomic chain with impurity will larger than that of the corresponding ideal atomic chain at certain value $\delta_{\mathrm{c}}$. The value, definitely, is called the \textit{critical mass ratio} throughout the paper.}}
\subsection{PBC}
\begin{figure}[!h] %[t]
\centerline{\includegraphics[width=9.0cm,height=5.0cm]{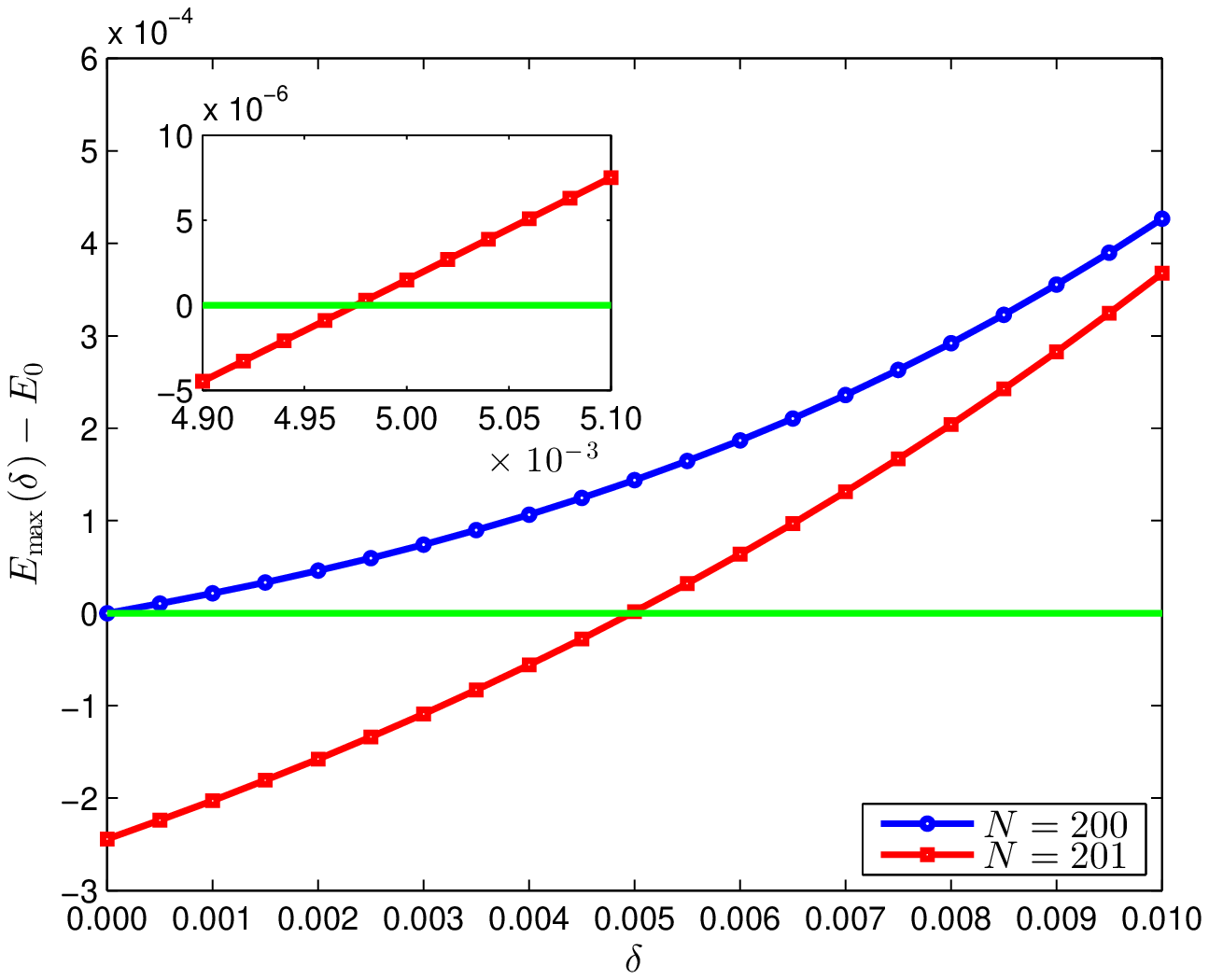}\hspace{4mm}}
%\centerline{\includegraphics[height=5.0cm,width=16.5cm]{fig1.eps}\hspace{4mm}}
%\begin{minipage}[b]{9cm}
\caption{The maximal vibration energy $E_{\max}(\delta)$~(with $E_0$=4) varies with the mass ratios $\delta$ for even~(blue) and odd~(red) number of total atoms. It can be distinguished that for $N=200/201$, the critical mass for the even case is zero, while for the odd case it satisfies the inequality $4.95\times10^{-3}<\delta_{\mathrm{c}}<5.00\times10^{-3}$~(see the insert), highly in accordance with the theoretical value $4.975\times10^{-3}$.}\label{FIG-CMvsEmax}
%\end{minipage}
\end{figure}

If the chain is PBC, only the mass of the impurity matters, and the typical matrix similar to \eref{matrixM} is
\begin{equation}\label{matrixMPBC}
M^{(1)}=
\left(
  \begin{array}{ccccccc}
    2-E   & -1      &        &               &           &           & -1       \\
    -1    & 2-E     &     -1 &               &           &           &          \\
          & \ddots  & \ddots & \ddots        &           &           &          \\
          &         & -1     &2-(1-\delta)E  &  -1       &           &          \\
          &         &        & \ddots        &  \ddots   &    \ddots &          \\
          &         &        &               &  -1       &    2-E    & -1       \\
    -1    &         &        &               &           & -1        & 2-E      \\
  \end{array}
\right).
\end{equation}
We use the The Molinari formula to calculate the determinant, where the properties of the trace are also considered. After some basic algebraic operators we arrive at
\begin{equation}\label{detMPBC-impurity}
\vert M^{(1)}\vert = 2\left(\cos N\tau-1\right)+2\delta\tan\frac{\tau}{2}\sin N\tau.
\end{equation}
In the spectra of the atoms, only one of them at most will be the local vibration energy, depending on the value of the mass ratio $\delta$. From \eref{dispersionRelation} we know that in the FBZ where $-\pi<\tau\leq\pi$, if $\tau$ is larger we can get the higher energy, and thus approach to the local vibration energy. It can be inferred that in the neighborhood of $\pi$ we should get the critical factor of mass ratio $\delta_{\mathrm{c}}$. Let $\tau=\pi+\varepsilon$ where $\varepsilon$ is a negative but infinitesimal number, we have
\begin{equation}
\cos N(\pi+\varepsilon)-1 + \delta_{\mathrm{c}}(1-\cos(\pi+\varepsilon))\frac{\sin N(\pi+\varepsilon)}{\sin (\pi+\varepsilon)}=0,
\end{equation}
and from which we get
\begin{equation}\label{deltacPBC}
\delta_{\mathrm{c}}=\frac{(-1)^{N+1}+1}{2N}=
   \cases
   {0,              &$N=2,4,\cdots$ \\
   \frac{1}{N},     &$N=1,3,\cdots$}.
\end{equation}
It came as a surprise that the critical mass ratio should show an \textit{odd-even effect}, {\color{blue}{which can also be verified numerically as illustrated in \fref{FIG-CMvsEmax}.}}

In summary, in the PBC case, if the number of the atoms in the chain is even, the condition for the emergence of the local vibration is $m'<m$, compared to $m'<m(1-1/N)$ for the odd case. However, in the limit where $N\rightarrow\infty$, the odd-even effect vanishes and the critical mass ratios are the same for the two cases as the bulk behaviors.

\subsection{OBC}
If the chain is OBC, we should specific the position of the impurity. If the impurity locates at the end of the chain~(referred as case $a$), say the first entry of the chain, the matrix is
\begin{equation}\label{matrixMOBC-A}
M_{a}^{(0)}=
\left(
  \begin{array}{ccccccc}
    1-(1-\delta)E   & -1      &        &               &           &           &          \\
    -1              & 2-E     &     -1 &               &           &           &          \\
                    & \ddots  & \ddots & \ddots        &           &           &          \\
                    &         & -1     &2-E            &  -1       &           &          \\
                    &         &        & \ddots        &  \ddots   &    \ddots &          \\
                    &         &        &               &  -1       &    2-E    & -1       \\
                    &         &        &               &           & -1        & 1-E      \\
  \end{array}
\right).
\end{equation}
The determinant of \eref{matrixMOBC-A} turns to be
\begin{equation}\label{detMOBCA-impurity}
\vert M_{a}^{(0)}\vert = -2\tan\frac{\tau}{2}\left[\sin N\tau-2\delta\cos\left(N\tau-\frac{\tau}{2}\right)\sin\frac{\tau}{2}\right]
\end{equation}
with the help of the Molinari formula, and the critical mass ratio
\begin{equation}\label{deltacOBC-A}
\delta_{\mathrm{c}}=\frac{N}{2N-1} \gtrsim \frac{1}{2}   %\lesssim
\end{equation}
can be obtained by the same method.

\begin{figure}[!h] %[t]
\centerline{\includegraphics[width=9.0cm,height=5.0cm]{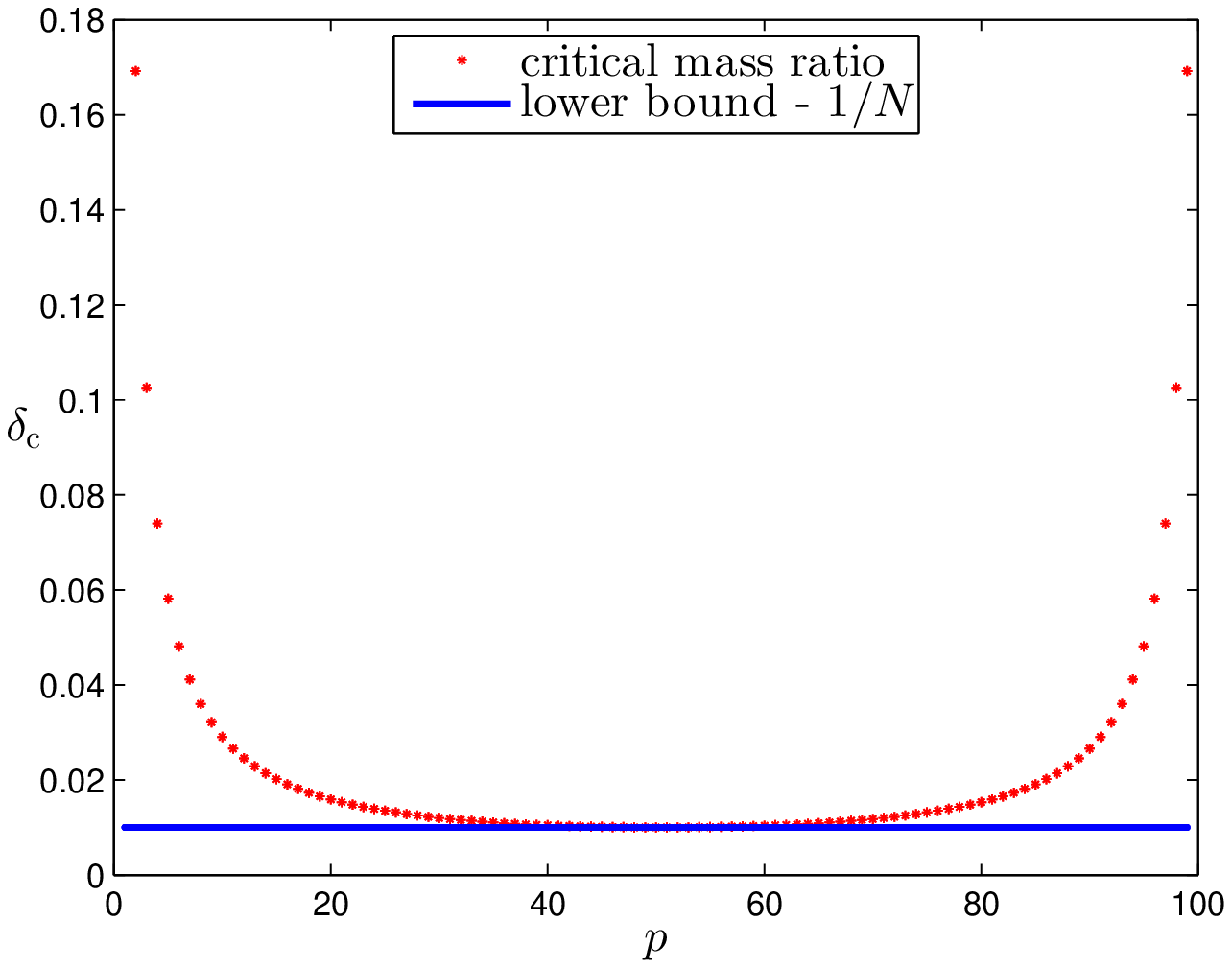}\hspace{4mm}}
%\centerline{\includegraphics[height=5.0cm,width=16.5cm]{fig1.eps}\hspace{4mm}}
%\begin{minipage}[b]{9cm}
\caption{The critical mass ratio $\delta_{\mathrm{c}}$  varies with the position $p$~($1<p<N$) of the impurity with $N=100$ under OBC.}\label{FIG-CMR}
%\end{minipage}
\end{figure}

If the impurity locates at the middle of the chain~(referred as case $b$), where the number of atoms at the left and right hand sides of the impurity are $N_l$ and $N_r$, respectively. The matrix is
\begin{equation}\label{matrixMOBC-B}
M_{b}^{(0)}=
\left(
  \begin{array}{ccccccc}
    1-E             & -1      &        &               &           &           &          \\
    -1              & 2-E     &     -1 &               &           &           &          \\
                    & \ddots  & \ddots & \ddots        &           &           &          \\
                    &         & -1     &2-(1-\delta)E  &  -1       &           &          \\
                    &         &        & \ddots        &  \ddots   &    \ddots &          \\
                    &         &        &               &  -1       &    2-E    & -1       \\
                    &         &        &               &           & -1        & 1-E      \\
  \end{array}
\right),
\end{equation}
and the corresponding determinant is
\begin{equation}\label{detMOBCB-impurity}
\vert M_{b}^{(0)}\vert = -2\tan\frac{\tau}{2}\left[\sin N\tau - \delta{\cal{F}}\right]
\end{equation}
where
\begin{equation}\label{detMOBC-BFx}
{\cal{F}} =\sin N\tau + \tan\frac{\tau}{2}\cos(N_l-N_r)\tau+\frac{\cos N\tau-\cos(N+1)\tau}{\sin\tau}
\end{equation}
with $N_l+N_r+1=N$. Similarly we have the critical mass ratio
\begin{equation}\label{deltacOBC-B}
\delta_{\mathrm{c}}=\frac{N}{(2N_l+1)(2N_r+1)} \geq \frac{1}{N}   %\lesssim
\end{equation}
where the equality holds unless the length of the chain is \textit{odd} and $N_l=N_r$. It is interesting to note that if $N_l=0$, which means that the impurity locates that the end of the chain, \eref{deltacOBC-B} should reduce to \eref{deltacOBC-A}, indicating a self-consistent calculation. \Fref{FIG-CMR} shows the critical mass ratio $\delta_{\mathrm{c}}$ varies with the position $p$ of the impurity under OBC with the total number $N=100$.

In summary, in the OBC situation, the critical mass ratio $\delta_{\mathrm{c}}$ severely depends on the position of the impurity. If the impurity locates at the end, the critical mass $\frac{N-1}{2N-1}m$ is sightly less than half of the perfect atom mass, and tends to $m/2$ for infinite chain. If the impurity locates at the middle, however, the critical mass ratio relies on its distance from the endpoint, and decreases dramatically from $\frac{N}{2N-1}$ to $\frac{1}{N}$ with the distance.

\section{vibration frequency and amplitude}\label{sec1Dmonatomicvibrationfrequency}
\subsection{vibration frequency}
It is necessary for us to investigate the frequency and amplitude of the vibration of impurity.  For the one impurity case, the frequency can be expressed as\cite{AD-2000,McCluskey-2000,YY-2014}
\begin{equation}\label{vibrationFrequency}
\omega_I = \frac{\omega_{\mathrm{m}}}{\sqrt{1-\delta_{f}^2}}
\end{equation}
where $\omega_{\mathrm{m}}=\sqrt{E_{\mathrm{m}}\beta/m}$ is the maximum frequency of the local vibration with $E_{\mathrm{m}}=4$ be the maximum energy of the corresponding perfect monatomic chain, and
\begin{equation}\label{deltap}
\delta_f \doteq \frac{m-fm'}{m}=1-f\frac{m'}{m}
\end{equation}
is a fitting parameter that depends on the boundary condition and the position of the impurity. In light of \eref{deltam}, \eref{vibrationFrequency} and \eref{deltap} the factor $f$ can be fitted as
\begin{equation}\label{pFitFun}
f=f_N(\delta)=\frac{1-\sqrt{1-\frac{E_\mathrm{m}}{E_I}}}{1-\delta}
\end{equation}
where $E_I=m{\omega_I}^2/\beta$ and is also the maximal energy of the total atoms.

\begin{table}[!h]
\centering
\caption{The local vibration energy $E_{I}$ of the impurity and the corresponding fitting factor $f$ for different boundary condition, length $N$ and mass ratio $\delta$.}\label{TAB-1}
\begin{tabular}{*{8}{c}}
\hline \hline
\multirow{2}*{\textrm{PBC}}
& \multicolumn{2}{c}{$N=100$}
& \multicolumn{2}{c}{$N=200$} \\\cline{2-5}
& $\delta=0.02$ & $\delta=0.20$
& $\delta=0.02$ & $\delta=0.20$\\
\hline \hline
$E_{I}$     &4.001706919850158		&4.166666666666721		&4.001602776644512		&4.166666666666705\\
\hline
$f    $     &0.999333635604874		&0.999999999999961		&0.999986390516793		&0.999999999999972\\
\hline \hline
\end{tabular}

\begin{tabular}{*{8}{c}}
\hline \hline
{\textrm{OBC}}
& \multicolumn{2}{c}{$N=100$}
& \multicolumn{2}{c}{$N=200$} \\\cline{2-5}
($a$)
& $\delta=0.02$ & $\delta=0.20$
& $\delta=0.02$ & $\delta=0.20$\\
\hline \hline
$E_{I}$     &4.001467394924789		&4.166666666666679		&4.001598481987127		&4.166666666666665\\
\hline
$f    $     &1.000867568004733		&0.999999999999991		&1.000013758106044		&1.000000000000001\\
\hline \hline
\end{tabular}

\begin{tabular}{*{8}{c}}
\hline \hline
{\textrm{OBC}}
& \multicolumn{2}{c}{$N=100$}
& \multicolumn{2}{c}{$N=200$} \\\cline{2-5}
($b$)
& $\delta=0.60$ & $\delta=0.80$
& $\delta=0.60$ & $\delta=0.80$\\
\hline \hline
$E_{I}$     &4.166666666666692		&6.250000000000008		&4.166666666666695		&6.250000000000003\\
\hline
$f    $     &1.999999999999964		&1.999999999999997		&1.999999999999959		&1.999999999999999\\
\hline \hline
\end{tabular}
\end{table}

\Tref{TAB-1} shows the local vibration energy $E_{I}$ of the impurity and the corresponding fitting factor $f$ for different boundary condition, length $N$ and mass ratio $\delta$. For any given boundary condition, the local vibration energy $E_{I}$ highly relies on the mass ratio, but the fitting factor $f$ only sightly varies. Thus, it can be concluded from \tref{TAB-1} that for PBC the factor $f=1$\cite{McCluskey-2000}, and for OBC the factor $f=1$ if the impurity locates at the middle while $f=2$ if the impurity locates at the end\cite{YY-2014}. Our numerical results suggest that splendid accuracy such as more that 15 decimal places can be achieved for the fitting factor $f$ as long as the length $N$ of the chain is long enough and/or the mass ratio $\delta$ is large enough.

\subsection{vibration amplitude}
The local vibration can be distinguished by the amplitudes of the atoms around the impurity. We will set the length of chain $N=200$ unless stated otherwise.

\begin{figure}[!h] %[t]
\centerline{\includegraphics[width=9.0cm,height=5.0cm]{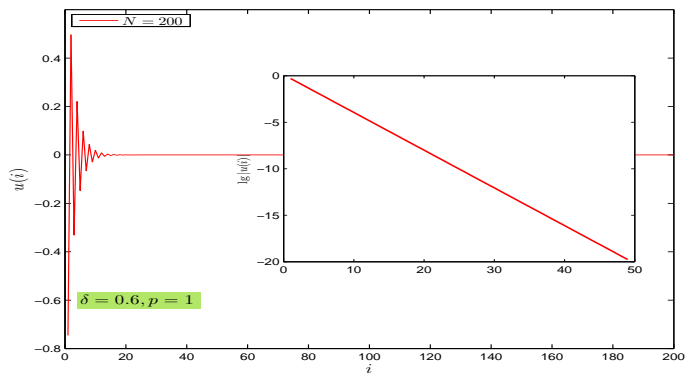}\hspace{4mm}}
%\centerline{\includegraphics[height=5.0cm,width=16.5cm]{fig1.eps}\hspace{4mm}}
%\begin{minipage}[b]{9cm}
\caption{The local vibration of impurity with mass ratio $\delta=0.60$ at $p=1$ of total length $N=200$. The critical factor is $0.5012531$, and the corresponding vibration frequency is 4.1666667.}\label{FIG-OBC-End}
%\end{minipage}
\end{figure}

\begin{figure}[!h] %[t]
\centerline{\includegraphics[width=9.0cm,height=5.0cm]{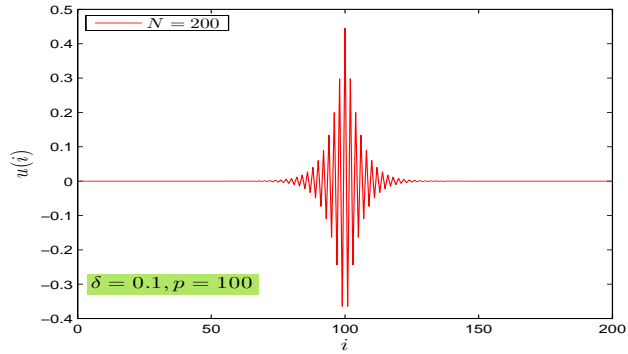}\hspace{4mm}}
%\centerline{\includegraphics[height=5.0cm,width=16.5cm]{fig1.eps}\hspace{4mm}}
%\begin{minipage}[b]{9cm}
\caption{The local vibration of impurity with mass ratio $\delta=0.60$ at $p=1$ of total length $N=200$. The critical factor is $0.0050001$, and the corresponding vibration frequency is 4.0404040.}\label{FIG-OBC-Mid}
%\end{minipage}
\end{figure}

\Fref{FIG-OBC-End} and \fref{FIG-OBC-Mid} show the occasions where the impurity locates at the end and middle, respectively, of the monatomic linear chain under OBC. In \fref{FIG-OBC-End} the amplitude of the leftmost impurity is the largest, while the amplitudes of the other atoms oscillate around zero and vanish exponentially along the chain\cite{YY-2014}, as can be seen from the insert of the \fref{FIG-OBC-End}. Actually, the amplitudes of the atoms can be analysized as follow. From the equation of motion of the atoms, we have
\begin{equation}\label{EOMamplitude}
   \cases{
   u_1-u_2 = (1-\delta)\frac{m\omega_I^2}{\beta}u_1,                            \\
   -u_{k-1}+2u_k-u_{k+1} = \frac{m\omega_I^2}{\beta}u_k, &$k=2,3,\cdots,N-1$    \\
   -u_{N-1}-u_N = \frac{m\omega_I^2}{\beta}u_N
   }
\end{equation}
where $\omega_I$ is the frequency of the impurity and can be obtained through \eref{vibrationFrequency} approximately.  After some basic algebraic operators we get the recurrence form $\frac{u_{k+1}}{u_k} = -\frac{1-\delta}{\delta}$, or equivalently,
\begin{equation}\label{EOMrecurUkU1}
u_k = u_1(-1)^{k-1}\left(\frac{1-\delta}{\delta}\right)^{k-1} = u_1(-1)^{k-1}e^{-\ln\left(\frac{\delta}{1-\delta}\right)(k-1)}.
\end{equation}
\Eref{EOMrecurUkU1} holds unless the decay ratio $t = 1/\ln\left(\frac{\delta}{1-\delta}\right)>0$, \textit{i.e.} $\delta>\frac{1}{2}$, which is in accordance with the conclusion in \eref{deltacOBC-A}.

In \fref{FIG-OBC-Mid} the amplitudes of the atoms around the impurity reduces gradually, and decay to zero or so at the left and right two endpoints. If the impurity locates at the very middle of the atomic chain, the amplitudes at the left and right sides of the impurity show somewhat symmetry, which is to say that the decay ratios $t_l$ and $t_r$ are the same. If not, the decay ratios $t_l$ and $t_r$ are different.

\begin{figure}[!h] %[t]
\centerline{\includegraphics[width=10.5cm,height=6.0cm]{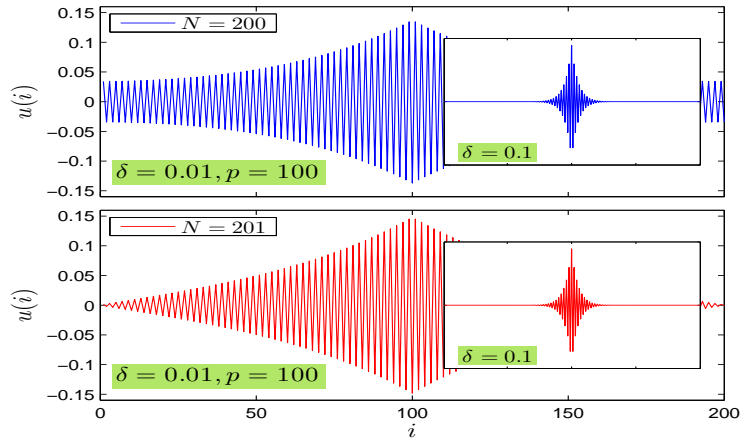}\hspace{4mm}}
%\centerline{\includegraphics[height=5.0cm,width=16.5cm]{fig1.eps}\hspace{4mm}}
%\begin{minipage}[b]{9cm}
\caption{The local vibration of impurity with $\delta=0.10$ at $p=100$ of total length $N=200$ and $201$, respectively. The critical mass ratios for the even and odd case are 0 and $0.004975$, and the corresponding vibration frequencies are 4.0004266 and 4.0003675, respectively.}\label{FIG-PBC}
%\end{minipage}
\end{figure}

\Fref{FIG-PBC} shows the amplitude behaviors of the monatomic linear chain with length $N=200$ and $N=201$ under PBC. For the odd case, the amplitudes will tend to zero or so with the increasing of the distance from the perfect atom to the impurity, and the amplitudes will become zero when the distance is $[N/2]$. For the even case, however, the amplitudes will only get small but not necessarily be zero for finite chain. With the increasing of the length $N$ or the mass ratio $\delta$, the amplitudes will become smaller and smaller, and the discrepancy between the even and odd cases will eliminate ultimately~(see the insert).

\section{Conclusion}\label{secConclusion}
In summary, we show an alternative derivation of the lattice vibration of one-dimensional monatomic linear chain under open boundary condition and periodic boundary condition through Molinari formula, and give the exact conditions for the emergence of the local vibration mode when one of the atoms is replaced by an impurity. Our derivation is different in that no coupled algebraic equations should be involved, and instead, we shall calculate the determinant of a tridiagonal matrix which can be handled by Molinari formula. Our method is more general than the above mentioned traditional method which can only settle the problem under periodic boundary condition.

Besides, we give the exact conditions for the emergence of the local vibration mode due to the impurity, which can be measured by the mass ratio $\delta=\frac{m-m'}{m}$ with $m'$ and $m$ be the mass of the perfect and impurity atoms, respectively. At periodic boundary condition, the critical mass ratio is $0$ or $\frac{1}{N}$, depending on whether the length $N$ of the chain is even or odd. At open boundary condition, the critical mass ratio is $\frac{N}{2N-1}$ if the impurity locates at the end of the chain, and is $\frac{N}{(2N_l+1)(2N_r+1)}$ with $N_l$ and $N_r$ be the number of atoms at the left and right hand sides of the impurity if the impurity locates at the middle of the chain. As to the vibration frequency, it can be expressed by $\omega_I = \frac{\omega_{\mathrm{m}}}{\sqrt{1-\delta_{f}^2}}$ where the fitting factor $f$ is 1 or 2 depending on the boundary condition and the position of the impurity along the atomic linear chain. The vibration amplitude of the atoms near the impurity decays exponentially with the distance between them.

\textit{Acknowledgements}
%\begin{acknowledge}
Qiang Luo would like to express his appreciation to Zhidan Wang for her careful reading of the manuscript, and to Ninghua Tong for some useful talk.
%\end{acknowledge}

\section*{Appendix A: Chebyshev polynomial for the determinant of tridiagonal matrix}\label{appA}
\setcounter{equation}{0}
\renewcommand{\theequation}{A.\arabic{equation}}

Let us utilize the Chebyshev polynomial\cite{cbsvpoly-url} to calculate the determinant of the tridiagonal matrix
\begin{equation}\label{tridiagmatrixFormab}
T=
\left(
  \begin{array}{ccccc}
    a     & b       &         &          &  b      \\
    b     & a       & b       &          &         \\
          & \ddots  & \ddots  & \ddots   &         \\
          &         & b       & a        &  b      \\
    b     &         &         & b        &  a      \\
  \end{array}
\right).
\end{equation}
$T$ is a symmetry circulant matrix, whose $m$-th eigenvalue is
\begin{equation}\label{chelambdam}
\lambda_m = a+b(\omega^m+\omega^{-m})=a+2b\cos{\frac{2\pi m}{n}}
\end{equation}
where $\omega=e^{2i\pi/n}$ is the $n$-th root of unity. The determinant, therefore, should be
\begin{equation}\label{chePBCDet}
\det{(T)} = \prod_{m=1}^{n}\lambda_m = \prod_{m=1}^{n}\left[a+2b\cos{\frac{2\pi m}{n}}\right] = b^n\prod_{m=1}^{n}\left[\frac{a}{b}+2\cos{\frac{2\pi m}{n}}\right]
\end{equation}
By virtue of the definition of the Chebyshev polynomial of the first kind $T_n$\cite{cbsvpoly-url}, \eref{chePBCDet} is readily to be
\begin{equation}\label{chePBCDet2}
\det{(T)} = 2(-b)^n\left[T_n\left(-\frac{a}{2b}\right)-1\right]
\end{equation}
Noticing that $T_n(-x)=(-1)^nT_n(x)$ and $T_n(\cos\tau)=\cos n\tau$, we finally reach at
\begin{equation}\label{chePBCDet3}
\det{(T)}=2b^n\left[(-1)^{n+1}+\cos n\tau \right].
\end{equation}

Similarly, the $m$-th eigenvalue of $\bar T$~(the two elements at the corner are zero) is
\begin{equation}\label{chebarlambdam}
\bar{\lambda}_m = a+b(\bar\omega^m+\bar\omega^{-m})=a+2b\cos{\frac{2\pi m}{n+1}}
\end{equation}
where $\bar\omega=e^{i\pi/(n+1)}$ is the $(2n+2)$-th root of unity. The determinant in this case is
\begin{equation}\label{cheOBCDet}
\det{(\bar T)} = 2b^nU_n\left(\frac{a}{2b}\right)
\end{equation}
where $U_n(x)$ is the Chebyshev polynomial of the second kind\cite{cbsvpoly-url}. Since $U_n(\cos\tau)=\frac{\sin (n+1)\tau}{\sin\tau}$, \eref{cheOBCDet} is nothing but
\begin{equation}\label{cheOBCDe2}
\det{(\bar T)}=b^n\frac{\sin(n+1)\tau}{\sin\tau}.
\end{equation}
%*******************************

\section*{Appendix B: Molinari formula for the determinant of tridiagonal matrix}\label{appB}
\setcounter{equation}{0}
\renewcommand{\theequation}{B.\arabic{equation}}

For any generalized tridiagonal matrix in the form of
\begin{equation}\label{tridiagmatrixForm}
T=
\left(
  \begin{array}{ccccc}
    a_1     & b_1     &             &               &  b_0           \\
    b_1     & a_2     & b_2         &               &                \\
            & \ddots  & \ddots      & \ddots        &                \\
            &         & b_{n-2}     &a_{n-1}        &  b_{n-1}       \\
    b_0     &         &             &b_{n-1}        &  a_{n}         \\
  \end{array}
\right)
\end{equation}
Molinari formula tells us that its determinant is
\begin{equation}\label{MoliDet}
\det{(T)} = 2(-1)^{n+1}\prod_{i=0}^{n-1}b_i + \mathrm{tr}\left[\prod_{k=1}^{n}
\left(
  \begin{array}{cc}
    a_k & -b_{k-1}^2 \\
    1   & 0          \\
  \end{array}
\right)
\right]
\end{equation}
Specially, \eref{tridiagmatrixForm} will be reduced to the ordinary tridiagonal matrix $\bar T$ when the entry at the corner is zero. We are not intended to repeat the lengthy proof here, while readers who are interested in the mathematical details can refer the original papers\cite{Molinari-1997,Molinari-2008}. Meaningfully, a typical example will be involved in this appendix to demonstrate that how to employ the Molinari formula to calculate the determinant, and the proof of a simplified version of the formula, \textit{i.e.}, $a_i=a$, $b_{i-1}=b$~($i=1,2,\cdots,n$)~(see \eref{tridiagmatrixFormab}) is provided as well for the benefit of completeness.

Let's introduce a $2\times2$ matrix
\begin{equation}\label{2time2matrix}
M=
\left(
  \begin{array}{cc}
    d & -1 \\
    1 & 0 \\
  \end{array}
\right)
\end{equation}
whose $n$-order power $M^{n}$ will be calculated subsequently. The eigenvalues of $M$ are $\lambda_{\pm}=e^{\pm i\tau}$, and the corresponding (unnormalized) eigenvectors are  $v_{\pm}=\left(e^{\pm i\tau},1\right)^{T}$ respectively with $\tau=\arccos(d/2)$. Therefore, the similarity transformation of $M$ is
\begin{equation}\label{Msimilar}
M=S\Lambda S^{-1}
\end{equation}
where $\Lambda=\mathrm{diag}\left(e^{+ i\tau},e^{- i\tau}\right)$, and the transformation matrix is
\begin{equation}\label{2time2matrixS}
S=
\left(
  \begin{array}{cc}
    e^{+ i\tau} & e^{- i\tau} \\
    1           & 1 \\
  \end{array}
\right).
\end{equation}
So, we have
\begin{equation}\label{nthMSlambdaS}
M^n=S\Lambda^nS^{-1}=\frac{1}{\sin\tau}
\left(
  \begin{array}{cc}
    \sin{(n+1)\tau} & -\sin{(n)\tau} \\
    \sin{(n)\tau}   & -\sin{(n-1)\tau} \\
  \end{array}
\right).
\end{equation}
When it comes to \eref{tridiagmatrixFormab}, Molinari formula tells us that
\begin{equation}\label{detabPBC}
\det{(T)}=2b^n\left[(-1)^{n+1}+\cos n\tau \right]
\end{equation}
and
\begin{equation}\label{detabOBC}
\det{(\bar T)}=b^n\frac{\sin(n+1)\tau}{\sin\tau},
\end{equation}
which are exactly the same as \eref{chePBCDet3} and \eref{cheOBCDe2}.

%%%%%%%%%%%%%%%%%%%%%%%%%%%%%%%%%%%%%%%%%%%%%%%%%%%%%%%%%%%%%%%%%%%%%%%%%%%%%%%%%%%%%%%%%
\section*{References}
%% ###################################################################
%\begin{small}

%\end{small}
%% ###################################################################


\begin{thebibliography}{99}
\bibitem{QLUO-Note} Strictly speaking, the so-called \textit{Brillouin zone} is illegal for the open boundary condition case in that the translation invariance property is not exist. So, the notation $q_0$ here should not be regard as the wavevector, but rather an index for energy spectrum.
\end{thebibliography}
\end{document}